\documentclass[fleqn,usenatbib]{mnras}

\usepackage{newtxtext,newtxmath}

\usepackage[T1]{fontenc}
\usepackage{ae,aecompl}

\usepackage{graphicx}	
\usepackage{amsmath}	
\usepackage{amssymb}	
\usepackage{textcomp}	

\title[Failed TDEs from the Galactic Center]{``Failed'' tidal disruption events and X-ray flares from the Galactic Center}

\author[Sacchi, A.]{
Andrea Sacchi$^{1,2,3}$,\thanks{E-mail: sacchi@arcetri.astro.it}
Giuseppe Lodato$^{3}$
\\
$^{1}$Dipartimento di Fisica e Astronomia, Universit\`a degli Studi di Firenze, Via G. Sansone 1, I-50019, Sesto Fiorentino, Italy\\
$^{2}$Dipartimento di Fisica, Universit\`a degli Studi di Pavia, Via Bassi 6, 27100, Pavia, Italy\\
$^{3}$Dipartimento di Fisica, Universit\`a degli Studi di Milano, Via Celoria 16, 20133, Milano, Italy\\
}

\date{Accepted XXX. Received YYY; in original form ZZZ}

\pubyear{2019}

\begin{document}
\label{firstpage}
\pagerange{\pageref{firstpage}--\pageref{lastpage}}
\maketitle

\begin{abstract}
The process of tidal disruption of stars by a supermassive black hole (SMBH) provides luminous UV and soft X-ray flares with peak luminosities of $\approx 10^{46}$ ergs/sec and duration of a few months. As part of a wider exploration of the effects of stellar rotation on the outcome of a TDE, we have performed hydrodynamical simulations of the disruption of a rotating star whose spin axis is opposite to the orbital axis. Such a retrograde rotation makes the star more resilient to tidal disruption, so that, even if its orbit reaches the formal tidal radius, it actually stays intact after the tidal encounter. However, the outer layers of the star are initially stripped away from the core, but then fall back onto the star itself, producing a newly formed accretion disc around the star. We estimate that the accretion rate onto the star would be strongly super-Eddington (for the star) and would result in an X-ray flare with luminosity of the order of $\approx 10^{40}$ ergs/sec and duration of a few months. We speculate that such events might be responsible for the known X-ray flares from Sgr A* in the recent past.
\end{abstract}

\begin{keywords}
black hole physics -- hydrodynamics -- galaxies: nuclei -- Galaxy: center
\end{keywords}



\section{Introduction}

Tidal disruption events (TDE) have been studied since the seventies of the past century, analytically \citep{lac82,ree88, phi89} and numerically \citep{lum82, lum83, eva89} and from the nineties also observationally \citep{kom99a}.

These events, firstly discovered as soft X-ray flares from otherwise quiescent galaxies \citep{bad96}, have been lately associated also with optical and UV transients \citep{gez12, kom12, kom15,hun17} and more recently with hard X-ray, gamma and radio emissions \citep{blo11, cen12, bro15,auc17,bla17}.

Interestingly, the very first theoretical investigations on this subject had our own Galactic Center in mind as a possible source of TDEs \citep{lac82}. However, no TDE has ever been observed from our galactic center (although some recent flare have been associated with the tidal disruption of asteroids \citealt{kos12}). On the other hand, there is evidence of past activity from Sgr A$^*$ in the form of rapid flare and variability, that can be inferred based on the echo that these flares produce on the surrounding medium. In particular, \citet{chu17} argue that a recent X-ray flare has occurred in the Galactic center. Thanks to X-ray observations of the molecular clouds near the Galactic Center they were able to infer that these clouds reflect emission originating $\approx 110$ years ago from the central black hole in our Galaxy. They conclude that the original flare had a luminosity of $\approx10^{41}$ erg/s in the 1-100 keV band, and lasted for a a period of less then a few years.

Theoretical investigations of the disruption of stars by supermassive black holes have concentrated mostly on the case where the star does not possess a spin. In this context, we know that the resulting stream structure and TDE features (such as its fallback curve) depend on several quantities: the internal structure of the star \citep{lod09,cou15}, the penetration parameter $\beta$ \citep{gui13}, defined as the ratio between the tidal radius $R_\textup{t}=(M_\textup{h}/M_\star)^{1/3}R_\star$ and the pericenter of the stellar orbit, and the black hole spin \citep{tej17}. 

While basic dynamical arguments readily show that a TDE induces a spin in the disrupted star (in the same sense of the orbital angular momentum), only a couple of works \citep{gol19, kag19} to date investigated the effect of stellar spin on the TDE characteristics (and only over a limited range of parameters). In this paper, we concentrate on one specific setup, in which the stellar rotation axis is antiparallel to the orbital angular momentum of the star, so that the spin is retrograde with respect to the orbit. We show that such a configuration can lead to a ``failed'' TDE, where a star with $\beta=1$ (which would be disrupted in the absence of spin) can survive the encounter with the SMBH and rather produce a fainter flare, whose features resemble the observed flares from Sgr A$^*$. 

This work is divided as follow: in Section 2 we will present the standard picture of TDEs and how an initial stellar rotation can suppress the disruption; in Section 3 we will discuss the numerical setup used in our simulations; finally in Section 4 there will be a discussion of our results and our conclusions. 

\section{Analytical estimates}
In this section we briefly summarize the standard analytical estimates on TDE dynamics, mainly following the approach of \citet{ree88}. We will also give a taste of the relation between tides and rotation and then we will show how an initial stellar rotation affects these estimates and how it can even suppress a TDE.
\subsection{The standard picture}
Let us consider a star, of radius $R_\star$ and mass $M_\star$, on a parabolic orbit around a black hole of mass $M_\textup{h}$. If the orbital pericenter is within the tidal radius: 
\begin{equation}
R_\textup{t}\simeq R_\star\left(\frac{M_\textup{h}}{M_\star}\right)^{1/3}\approx 0.5\,\textup{AU}\left(\frac{R_\star}{R_\odot}\right)\left(\frac{M_\textup{h}}{10^6\,M_\odot}\right)^{1/3}\left(\frac{M_\star}{M_\odot}\right)^{-1/3},
\end{equation}
the distance at which tidal forces equal the stellar self-gravity, the star will be destroyed by the black hole tidal forces. After the disruption roughly half of the stellar debris will be launched on hyperbolic orbits, while the remaining half will be bound to the black hole. These bound debris will form an accretion disc around the black hole \citep{bon16} and eventually they will get accreted.

In the simplest scenario it is possible to estimate the lightcurve of the event through the mass return rate at pericenter: the two quantities will be proportional under the assumption that the circularization and accretion time are considerably shorter than the orbital period of the debris. While the validity of the previous assumptions is currently under debate (the disc might not form efficiently, \citealt{svi17}, and even if the disc forms the single wavelength lightcurve might deviate significantly from the bolometric behaviour, \citealt{lod11}), there is some empirical evidence that the optical luminosity does follow approximately the behaviour predicted analytically, hence giving credit that it might be related directly to the fallback process \citep{lod12}.
In this picture, the luminosity of the event will scale as the rate at which the stellar debris fall back onto the black hole, which is easily computed from Kepler's third law, assuming the debris to follow Keplerian orbits after disruption. This gives the standard result
\begin{equation}
L\propto\frac{\textup{d}M}{\textup{d}t}=\frac{(2\pi GM_\textup{h})^{2/3}}{3}\frac{\textup{d}M}{\textup{d}E}t^{-5/3}.
\end{equation}
The energy spread $\Delta E$ of the debris right before the disruption is determined only by how deep in the potential well the debris are. It is therefore possible to estimate the energy spread in a straightforward way as
\begin{equation}
\Delta E=\frac{\textup{d}E}{\textup{d}r}\bigg|_{R_\textup{p}}\Delta r=\frac{GM_\textup{h}}{R_\textup{p}^2}R_\star.
\end{equation}
Defining the minimum return time 
\begin{equation}
t_\textup{min}=\frac{2\pi GM_\textup{h}}{(2\Delta E)^{3/2}},
\end{equation}
as the time it takes for the first debris to come back at pericenter and the fallback peak rate 
\begin{equation}
\dot M_\textup{p}=\frac{M_\star}{3t_\textup{min}},
\end{equation}
it is possible to write down the fallback rate in a compact way as follow:
\begin{equation}
\frac{\textup{d}M}{\textup{d}t}=\dot M_\textup{p}\left(\frac t{t_\textup{min}}\right)^{-5/3}
\end{equation}
This result and the $-5/3$ power-law time decay has been traditionally considered the signature of TDE (even if later works, i.e. \citealt{lod09} showed that this time dependency is only reached at later times and it is strongly affected by the stellar internal structure).

\subsection{The effect of tidal forces on stellar rotation}
It is interesting to notice that even if the incoming star does not posses any initial proper rotation, the tidal forces induce one. 

\begin{figure}
	\includegraphics[width=0.7\columnwidth]{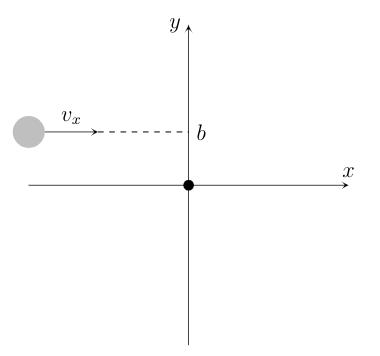}
    \caption{Schematic view of the geometry of the example proposed.}
    \label{fig:1a}
\end{figure}

To heuristically derive this fact we can imagine the star as moving on a $x-y$ plane with positive velocity $v_x$ towards the source of gravitational field (a mass $M_\textup{h}$ placed in the origin of the coordinates system) and impact parameter $b$ (figure \ref{fig:1a}).
In the impulse approximation the star will perceive the gravity source upon reaching the $y$ axis, at that moment it will feel an acceleration directed along the negative direction of the $y$ axis of magnitude $GM/b^2$ for a time $2b/v_x$: this will generate a velocity along $y$ given by
\begin{equation}
v_y=-\frac{GM}{b^2}\frac{2b}{v_x}=-\frac{2GM}{bv_x}.
\end{equation}
Since the orbital energy must be conserved during the passage, there will also be a change in the $v_x$ component:
\begin{equation}
\Delta v_x\approx -\frac{v_y^2}{2v_x}=-\frac{2G^2M^2}{b^2v_x^3},
\end{equation}
where given the impulse approximation (i.e. $\Delta v_x/v_x\ll1$) we neglected the square term in $\Delta v_x$.
One can see that this change in the velocity depends on the closeness with respect to the gravity field source $b$, so that material that is closest to the hole will experience a stronger reduction in horizontal velocity: the star will thus gain a spin, prograde with respect to the orbital motion, that we can estimate as
\begin{equation}
\omega\approx\frac{\textup{d}\Delta v_x}{\textup{d}b}=\frac{4G^2M^2}{b^3v_x^3}=\sqrt{\frac{2GM_\star}{R_\star^3}},
\end{equation}
where we assumed a parabolic orbit for the star with pericenter equal to the tidal radius.
We can compare this value with the break-up velocity of the star: defined as the velocity at which the centrifugal force overcome the stellar self-gravity:
\begin{equation}
\omega_\textup{b}=\sqrt{\frac{GM_\star}{R_\star^3}}.
\end{equation} 
This simple calculation, although approximated, shows therefore that the induced rotation exceed easily the break-up velocity of the star: tidal forces acts inducing a strong rotation in the approaching star.

\subsection{The effect of stellar rotation on tidal disruption}
It is possible to analyze the effects of an initial stellar rotation on a TDE using the same approximations of the picture presented above. The main difference will lie in the energy spread calculation: with respect to the previous case of a non-spinning star, now also the kinetic energy will depend on the position of the debris.

Let us assume the stellar rotation to be rigid: the angular velocity, $\omega$, is constant within the star and parametrized by the dimensionless quantity
\begin{equation}
\alpha=\frac{\omega}{\omega_\textup{b}}.
\end{equation}
Considering the rotation axis perpendicular to the orbital plane, the kinetic energy of the debris will be
\begin{equation}
E_\textup{k}=\frac12\left[v_\textup{o}+\alpha\omega_\textup{b}(r-R_\textup{p})\right]^2,
\end{equation}
where 
\begin{equation}
v_\textup{o}=\sqrt{\frac{2GM_\textup{h}}{R_p}}
\end{equation}
is the (parabolic) orbital velocity of the debris. Positive and negative values of $\alpha$ account for prograde and retrograde rotations respectively.

Taking into account the stellar rotation, the energy spread will have therefore an extra term:
\begin{equation}
\Delta E=\frac{\textup{d}}{\textup{d}r}\bigg|_{R_\textup{p}}(E_\textup{k}+E_\textup{p})=\frac{GM_\textup{h}}{R_\textup{p}^2}R_\star(1+\sqrt{2}\alpha).
\end{equation}
Therefore, positive values of $\alpha$ (prograde rotation) tend to increase the energy spread and make the star more prone to disruption, while negative $\alpha$ result in a reduction of the energy spread, leading to a longer $t_\textup{min}$ and a fainter TDE. Retrograde rotation makes a star more resilient to tidal disruption. In particular, there is a value of stellar rotation, $\alpha=-1/\sqrt{2}$, for which the energy spread is null: a star with such initial rotation will not be disrupted and thus in this case the TDE will be suppressed.

Obviously, given that tidal disruption induces spins comparable to break up, in order to have some effect on the TDE the initial stellar spin needs to be similarly high, close to break up. This is a quite considerable magnitude of stellar rotation: the sun, for example, spins at only  2\textperthousand\, of its break-up velocity. However the star in Galactic Centers should be spinning at an higher rate due to tidal interactions and spin up \citep{ale01}.

\section{Numerical simulation}
The model presented above predicts that retrograde rotation should suppress a TDE. To better investigate this phenomenon and to find out what would happen to the non-disrupted star we test the analytical predictions against numerical simulations.

The simulation was performed using the Smoothed Particle Hydrodynamic (SPH) code PHANTOM \citep{pri17}. This choice is due to the fact that in SPH the resolution "follows the mass" and in our simulation the majority of space is empty. 
The star was modeled as a sphere with a polytropic internal structure (adiabatic index $\gamma=5/3$). The mass and radius of the star are respectively $1\,M_\odot$ and $1\,R_\odot$, that are also our code units.  The velocity unit is $v_0=\sqrt{GM_\odot/R_\odot}\approx4.37\cdot10^7$ cm/s, the energy unit $E_0=v_0^2M_\odot\approx3.79\cdot10^{48}$ erg, the density unit $\rho_0=M_\odot/R_\odot^3\approx5.90$ g/cm$^3$ and finally the time unit for the system is $t_0=R_\odot/v_0\approx1590$ s.

In all our simulations we used approximately $5\cdot10^5$ SPH particles, however we tested the convergence of our results running simulations with $2.5\cdot10^5$ and $10^5$ particles and finding no appreciable differences.

In the following subsection we will illustrate how we set our stars into rotation in our broad study of the role of stellar rotation on the behavior of TDEs, as this is a quite delicate phase, and then focus on the specific setup we discuss in this paper.

\subsection{The relaxation process}
The sphere is initially relaxed in the absence of rotation or tidal field of the black hole in order to remove any noise in the initial random displacement of the SPH particles. During this phase, we apply a velocity damping, in order to reach equilibrium. We then check that the density profile does follow the expectations of a polytropic sphere (this is similar to what was done in \citealt{lod09}). 

\begin{figure}
\includegraphics[width=\columnwidth]{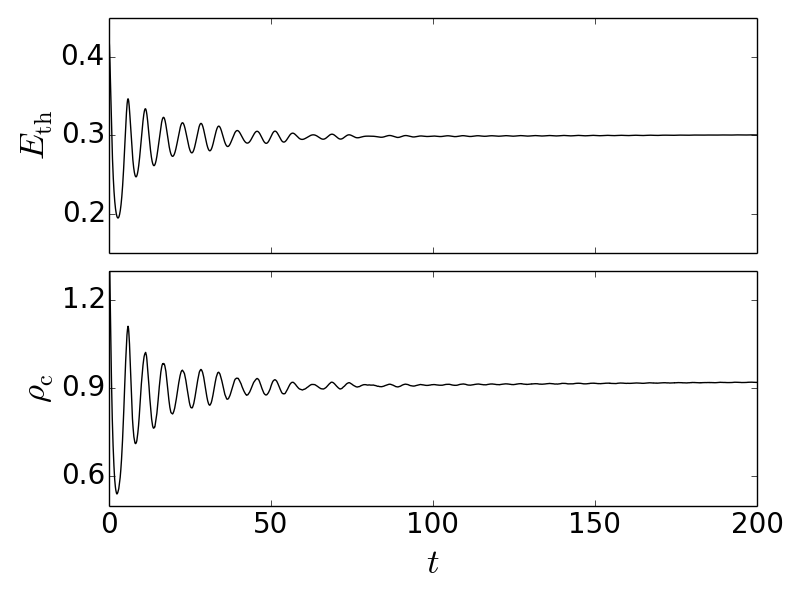}
    \caption{Thermal energy (top panel) and central density (bottom panel) during the relaxation process after the rigid rotation has been imposed as a function of time. After some initial strong oscillations the star reach equilibrium after t=200 (in code units).}
    \label{fig:relax}
\end{figure}

After this first relaxation a rigid rotation is imposed in the inertial frame of the star and then it is relaxed once more until it reaches equilibrium. During this phase, no additional velocity damping is applied, as it would also damp rotation, so relaxation only occurs through internal dissipative processes in SPH, related to artificial viscosity. To check that we reach a new equilibrium, we monitor the oscillations in the star's central density and total thermal energy.
Figure \ref{fig:relax} shows the thermal energy and the central density of the sphere during the relaxation process. They both share a similar behavior: at the beginning strong oscillations get progressively damped until they disappear and the star is considered at equilibrium and the relaxation process is complete.  

\begin{figure}
	\includegraphics[width=\columnwidth]{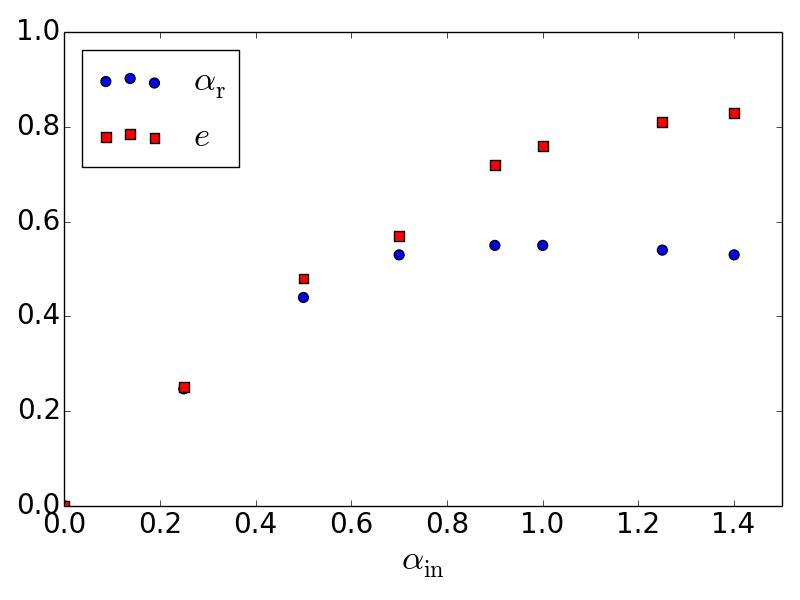}
    \caption{The angular velocity at the end of the relaxation process $\alpha_\textup{r}$ and polar eccentricity $e$ as functions of the initial angular velocity of the star $\alpha_\textup{in}$. After a first common linear growths the first tends to saturate while the second keeps increasing.}
    \label{fig:alpha}
\end{figure}

It is interesting to notice that as we increase the imposed rotation (hereinafter indicated with $\alpha_\textup{in}$), the rotation at the end of the relaxation process ($\alpha_\textup{r}$), does not grow linearly with it. While for $\alpha_\textup{in}\lesssim 0.5$ we have $\alpha_\textup{r}\approx \alpha_\textup{in}$, deviations start appearing for larger initial rotation rates and the final relaxed rate saturates to $\alpha_\textup{r}\approx 0.55$, regardless of the initial amount of rotation imposed.
This is due to the fact that as $\alpha_\textup{in}$ grows the star expands, conserving the total angular momentum but decreasing the rotation angular velocity. 
Figure \ref{fig:alpha} shows $\alpha_\textup{r}$ and the polar eccentricity $e=\sqrt{a^2+b^2}/a$ (where $a$ and $b$, the major and minor semi-axis respectively, are calculated setting a cut-off on the density equal to $10^{-2}$ code units) as functions of $\alpha_\textup{in}$ in our simulations. The parameter $\alpha_\textup{r}$ is calculated as the average on the angular velocity of the particles within the bulk of the star (that is, the particles rigidally rotating around the spin axis). It is notable that after an initial common linear growths, $\alpha_\textup{r}$ saturates around a value of $\alpha_\textup{r}\approx0.55$, while the polar eccentricity keeps growing although at a lower rate: this is due to the fact that the star flattens as $\alpha_\textup{in}$ increase.

Given the presence of the saturation, stars with the same $\alpha_\textup{r}$ could come from different $\alpha_\textup{in}$ and hence they have different internal structure and a different total angular momentum.

\begin{figure}
	\includegraphics[width=\columnwidth]{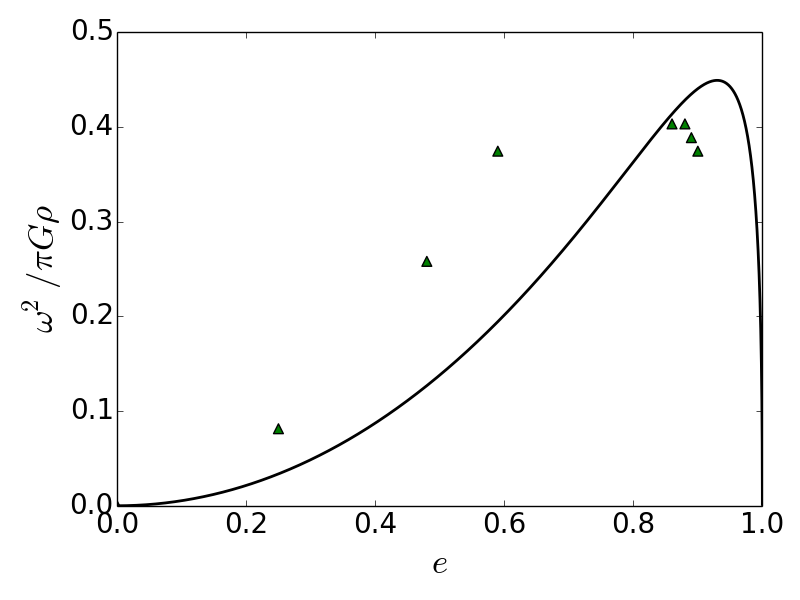}
    \caption{Normalized squared angular velocity as a function of the polar eccentricity of our simulation (colored triangles) compared with the McLaurin analytical series (solid black line).}
    \label{fig:mclaurin}
\end{figure}

This behavior follows the analytical predictions on the polar eccentricity and angular velocity for rotating spheroids, in particular the McLaurin series \citep{cha87}.
Figure \ref{fig:mclaurin} shows the normalized squared angular velocity of our simulated rotating spheres as a function of the polar eccentricity (green triangles), superimposed to the McLaurin analytical series (solid black line). The agreement is good, considering that the McLaurin series is computed for uniform density sphere while we deal with a polytropic internal structure.

\subsection{Numerical setup}
For our simulation the sphere is relaxed with a value of $\alpha_\textup{in}$ equal to $1$ and a subsequent $\alpha_\textup{r}=0.55$.

\begin{figure}
	\includegraphics[width=\columnwidth]{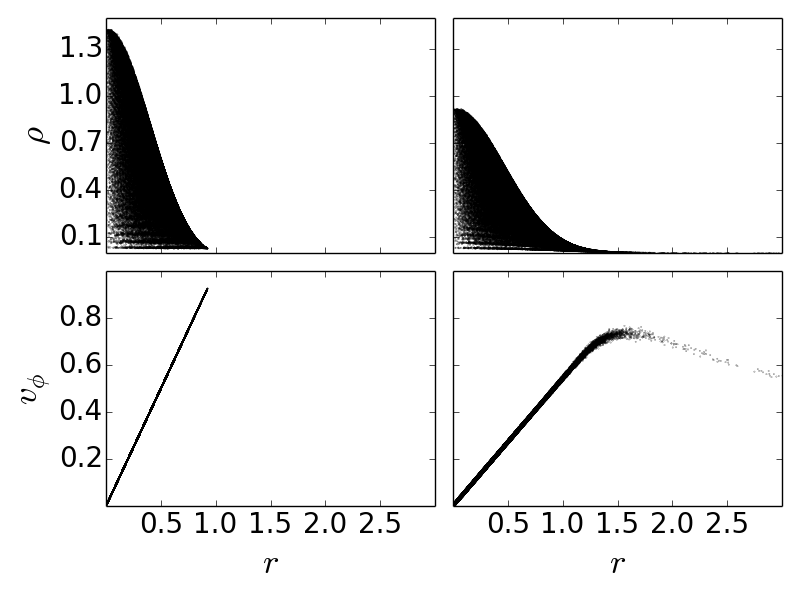}
    \caption{Comparison between the stellar properties before (left side) and after (right side) the relaxation process. The top panels show the density and the bottom ones the azimuthal velocity as functions of the distance from the rotation axis.}
    \label{fig:1}
\end{figure}

Figure \ref{fig:1} shows the density (top panel) and azimuthal velocity (bottom panel) of our sphere as functions of the distance from the rotation axis before (left panel) and after (right panel) the relaxation process. It is interesting to note that after the relaxation the bulk of the star keeps rotating rigidly, although at a reduced rate (see above). Also, a tail of particles with approximately Keplerian velocity profile is formed, with a negligible total mass.

The relaxed sphere is then injected into a parabolic orbit around a black hole (modeled as a Keplerian potential) with pericenter equal to the non-spinning tidal radius (impact parameter $\beta=R_\textup{t}/R_\textup{p}=1$) at a distance of 3 tidal radii. For such a configuration, a non-spinning star would be completely disrupted. The star is set to have a retrograde rotation with spin axis perpendicular to the orbital plane.

\section{Results}

\begin{figure}
\includegraphics[width=\columnwidth]{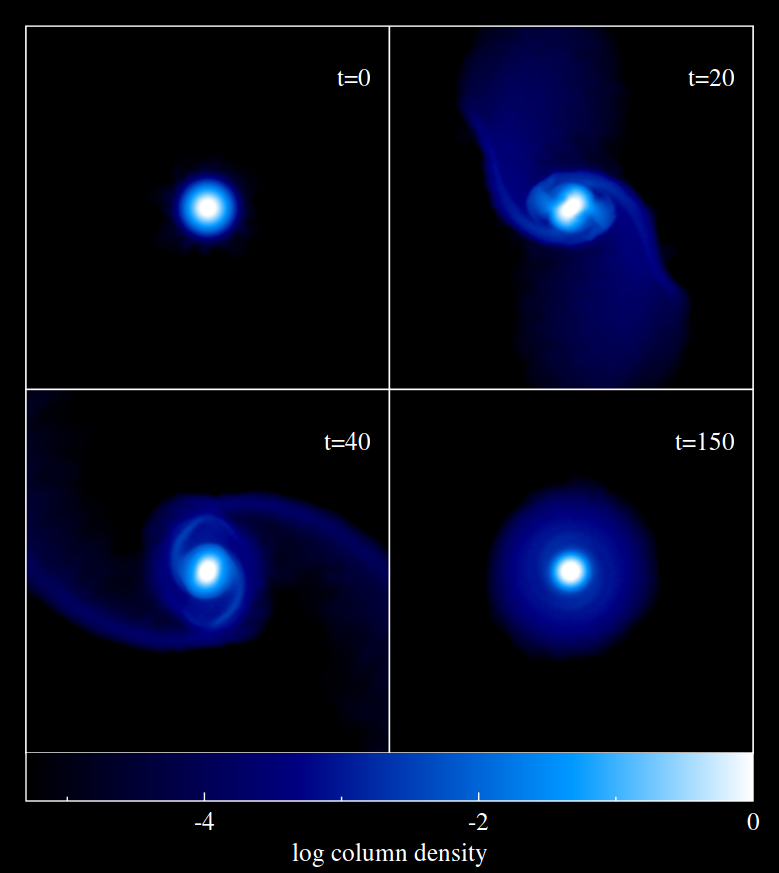}
    \caption{Projected density of the star, each panel shows a snapshot taken at a different time, as indicated in the figure in code units. In physical units the snapshots refer roughly to $t=0$, 9 hours, 18 hours and 3 days after the pericenter passage, respectively. The star is seen face-on.}
    \label{fig:2}
\end{figure}
\begin{figure}
\includegraphics[width=\columnwidth]{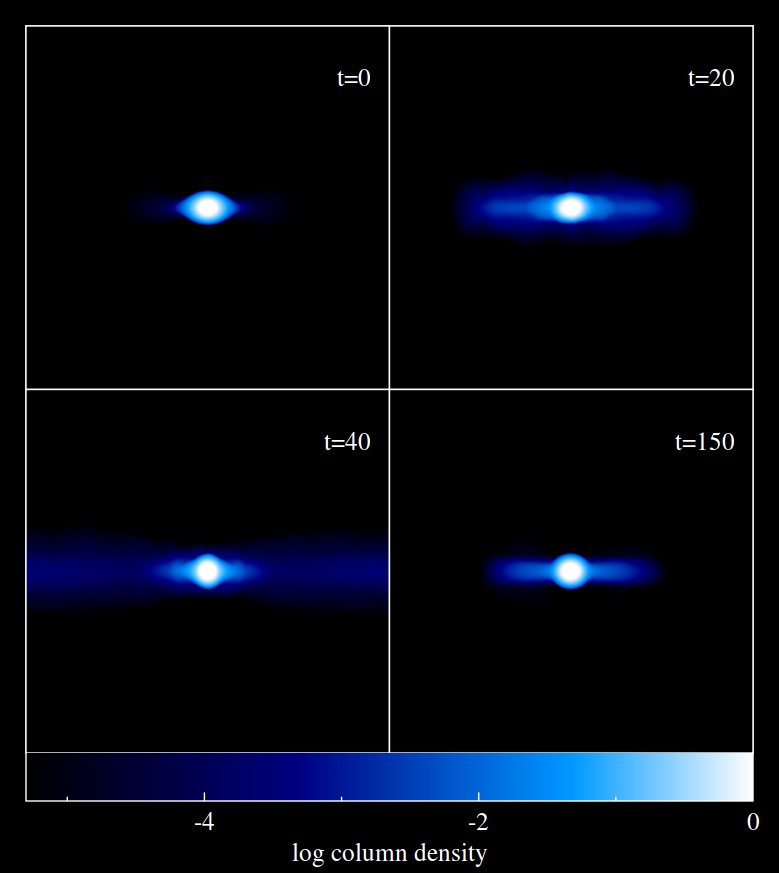}
    \caption{Projected density of the star, each panel shows a snapshot taken at a different time, as indicated in the figure in code units. In physical units the snapshots refer roughly to $t=0$, 9 hours, 18 hours and 3 days after the pericenter passage, respectively. The star is seen edge-on.}
    \label{fig:3}
\end{figure}

Our simulation shows that, even though the star formally reaches its tidal radius, it does not get destroyed: as predicted by our simple analytical model, an initial retrograde stellar rotation can indeed suppress the tidal disruption of a star. Nevertheless, the black hole tides are able to rip off some material from the star, however these debris do not accrete onto the black hole, indeed, at the end of the simulation, the mass loss suffered by the star is less then the $0.5\%$ of its initial mass: their fate is to be reaccreted onto the survival stellar core. In doing so they form an accretion disc with a Keplerian velocity profile.

Figure \ref{fig:2} and \ref{fig:3} show four snapshots of our simulation taken at the time $t=0$, $t=20$, $t=40$ and $t=150$ in our code units. The star is seen from above and from the side respectively and the rendering shows the projected density in code units. It is possible to see how the stellar material gets ripped off but eventually it falls back onto the stellar core as described above. In the last snapshot, taken at the end of our simulation, it is possible to recognize a disc.

\begin{figure}
\includegraphics[width=\columnwidth]{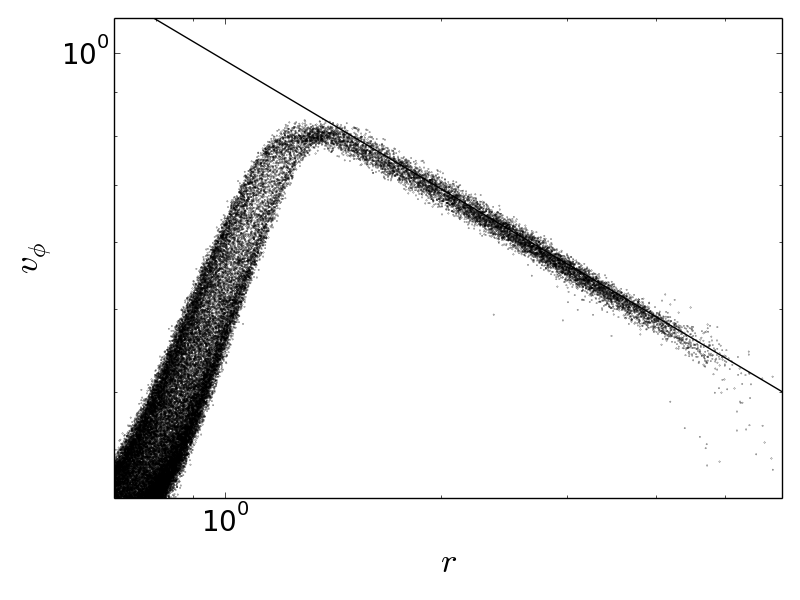}
    \caption{Velocity profile of the disc: azimuthal velocity as a function of the distance from the stellar core. The solid line is the $r^{1/2}$ behavior expected for a Keplerian disc.}
    \label{fig:4}
\end{figure}

Figure \ref{fig:4} shows the velocity profile of the stellar remnant at the end of the simulation. The logarithmic plot shows the azimuthal velocity as a function of the distance from the rotation axis in the stellar center of mass frame. The plot shows how the stellar core is in rigid rotation, while the particles surrounding the core have a Keplerian velocity profile (indicated by the solid line).

\subsection{Observational prospects}
At the end of our simulation we thus produce a thick accretion disc around a stellar core. The disc inner and outer radius (estimated from the velocity profile considering part of the disc only the particles with Keplerian velocity) are respectively $R_\textup{i}\approx 1.36\,R_\odot$ and $R_\textup{o}\approx 4.69\,R_\odot$, with a thickness of $H\approx 1.46\,R_\odot$. The mass of the disc amounts to $M_\textup{d}=3.18\cdot10^{-2}\,M_\odot$.

To infer the evolution of the disc we can compute the viscous time at the disc outer edge
\begin{equation}
t_\nu=\frac{R_\textup{o}^2}{\alpha_\textup{SS}H^2\Omega}\simeq1.67\cdot10^6\,\left(\frac{0.1}{\alpha_\textup{SS}}\right)\textup{sec}\approx0.05\,\left(\frac{0.1}{\alpha_\textup{SS}}\right)\textup{yr},
\end{equation}
where $\Omega$ is the angular velocity at the outer edge of the disc, given by
\begin{equation}
\Omega=\sqrt{\frac{GM_\textup{s}}{R_\textup{o}^3}}\simeq6.17\cdot10^{-5}\,\textup{sec}^{-1}.
\end{equation}
We also assume that the disc evolves viscously, with a \citet{sha73} parameter $\alpha_\textup{SS}$.
A rough estimate for the accretion rate can then be given as
\begin{equation}
\dot{M}_\textup{s}\approx \frac{M_\textup{d}}{t_{\nu}}\approx 0.63\left(\frac{0.1}{\alpha_\textup{SS}}\right)^{-1}M_{\odot}/\rm{yr}.
\end{equation}

Given these quantities the luminosity of this kind of event should amount to
\begin{equation}
L\approx \frac{GM_\textup{s}\dot{M}_\textup{s}}{2R_\odot}=2.67\cdot10^{40}\,\textup{erg/s}\left(\frac{\alpha_\textup{SS}}{0.1}\right)\approx200\,L_\textup{Edd}\left(\frac{\alpha_\textup{SS}}{0.1}\right),
\end{equation}
where the Eddington luminosity $L_\textup{Edd}\approx 1.26\cdot10^{38}$ erg/s is referred to the star.
The \citet{sha73} viscosity parameter, $\alpha_\textup{SS}$, in such a newly formed disc is not easy to estimate, given that we expect the magnetic field threading the disc to be weak and the magneto-rotational instability not to be effective \citep{bug18,nea18}. We have scaled it to $\alpha_\textup{SS}=0.1$ in the estimates above, but even if it was lower by two orders of magnitude, we should still expect an accretion event above Eddington for the star. 

At those luminosities, the expected temperature at the disc inner edge is $\approx 3.5\cdot10^5$ K, and thus, we expect such a hyper-accreting star to have significant emission in the x-rays band (young stars accreting at rates up to six order of magnitude smaller then our case have considerable emission in this band \citealt{fei07}), even though a smaller value of $\alpha_\textup{SS}$ could move it towards the UV-optical band.

The characteristics described above make our event compatible with the one described by \citet{chu17}: a bright x-rays flare with a duration no longer that a few years.

\subsection{Further stellar evolution}
At the end of our simulation the energy of the center of mass of the star is negative. This means that during the pericenter passage, the star gets captured by the black hole. 

From the energy of the center of mass of the star we calculated the orbital period of the newly bound star to be roughly one hundred years and its apocenter to be placed at $\approx4000$ au.

Furthermore from the $z$ component of the vorticity we can infer that the tidal forces are able, during the pericenter passage, to completely reverse the spin of the star. 

\begin{figure}
\includegraphics[width=\columnwidth]{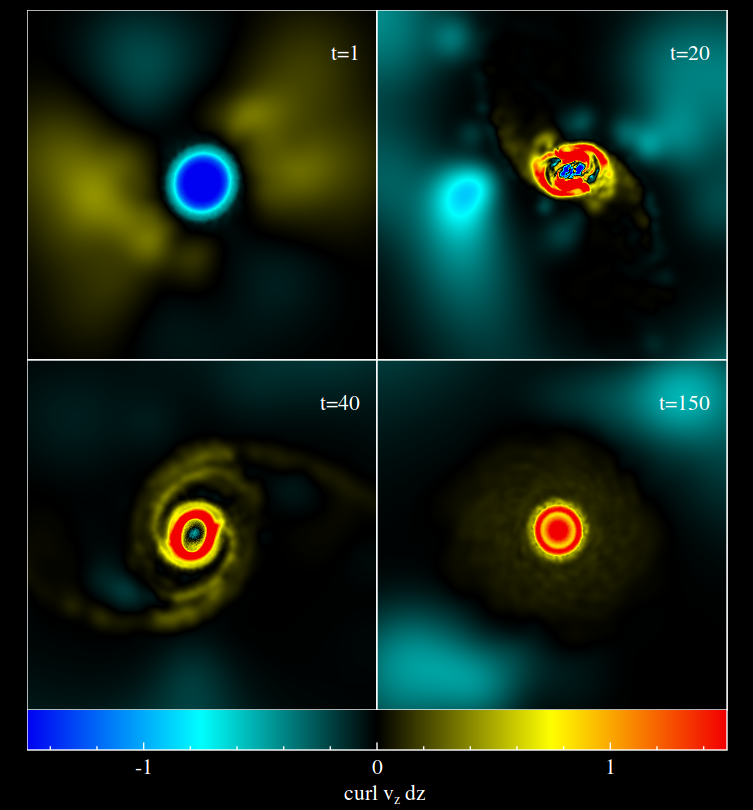}
    \caption{Each panel shows the $z$ component of the velocity at different times, as indicated in the figure in code units. In physical units the snapshots refer roughly to $t=0$, 9 hours, 18 hours and 3 days after the pericenter passage, respectively. The star is seen face-on.}
    \label{fig:5}
\end{figure}

Figure \ref{fig:5} shows the $z$ component of the vorticity at different times during the simulation. It is possible to notice how at the beginning the $z$ component is negative, i.e. the rotation is retrograde, while at the end it is positive: the rotation of the star is therefore prograde with the orbital motion with. Furthermore the rotation of the stellar core is rigid and with the remarkable magnitude of $\alpha\approx0.3$.

Taking into account that the star, after the passage, gets bound to the black hole and that its spin axis gets completely reversed, we can predict that roughly a century after the "failed" TDE, the star will approach again the black hole, this time however with a prograde and considerable spin, thus originating an enhanced TDE.

\section{Dependence on the orientation of the stellar spin}

The simulation described in section 4 above is referred to the particular case in which the stellar spin axis is exactly anti-parallel to the orbital angular momentum. 

As shown by \citet{gol19} the dependence of the features of a tidal disruption event on the orientation of the spin axis can be strong, and we therefore ran a few simulations with the stellar spin axis slightly misaligned with respect to the configuration considered until now.

As expected, we find that the formation of the disc of debris is inhibited as the deviation from the totally retrograde rotation regime grows. Although a disc forms, it is less massive than the perpendicular case, becasue more debris are ejected rather than re-accreted.

\begin{figure}
\includegraphics[width=\columnwidth]{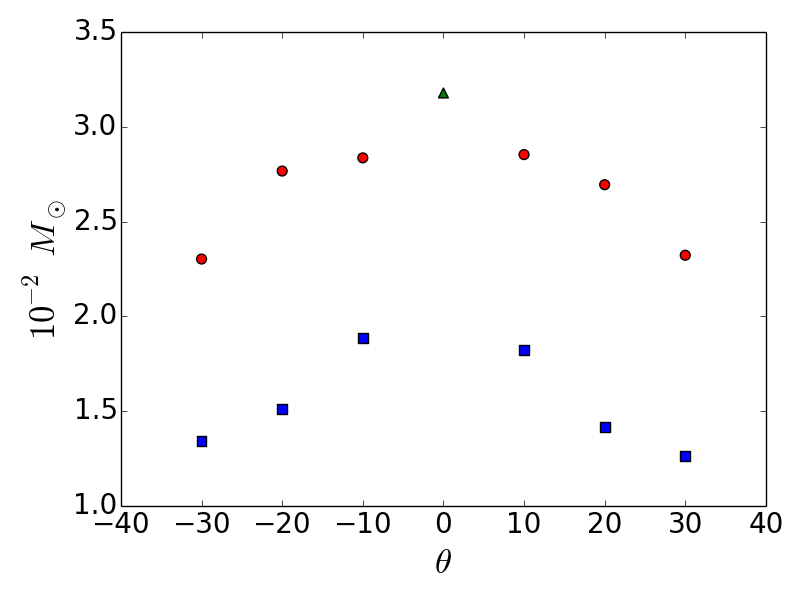}
    \caption{The mass of the disc at the end of our simulations, the green triangle is the completely retrograde case while the blue squares and red circles indicates tilting around $y$ and $x$ axis respectively.}
    \label{fig:disc_mass}
\end{figure}

Figure \ref{fig:disc_mass} shows the mass of the disc as a function of the tilt angle with respect to the retrograde regime. The green triangle represents the mass of the disc in the retrograde case, the blue squares a tilting along the $y$ axis while the red circles a tilting around the $x$ axis.

While the behaviour of the system is roughly symmetric to whether the tilting is right-handed (positive values of the angle) or left-handed (negative values), a tilting around $y$ clearly suppress the disc formation more aggressively than one around $x$. However even for a significant tilting of $30^\circ$ a disc still forms and its mass has dropped only by a factor $\approx 2.5$ for the $y$-axis tilting and $\approx 1.3$ if the tilting is performed around the $x$-axis.

\section{Conclusions}

In this paper we discuss the possibility of a "failed" tidal disruption event at the origin of a one hundred years old X-ray flare from SGR A$^*$. 

Through analytical calculation we find that a sufficiently strong stellar retrograde rotation could suppress its disruption during a black hole pericenter passage. We then performed a numerical simulation with the SPH code PHANTOM in order to validate our predictions. 

We discovered that even if the star does not get destroyed by the black hole tidal forces and thus no debris get accreted by the hole, significant electromagnetic emission can arise due to the formation of an accretion disc around the stellar core itself. This disc is made out of the material partially ripped off the star during the black hole flyby.

Through considerations about the size, mass and evolution  of this newly born disc we can infer that this event should produce a luminosity of $\approx3\cdot10^{40}$ erg/s on a period of few months. This makes this event compatible with the one described by \citet{chu17}.

Furthermore, as the star orbits the black hole, it gets captured by it and is placed on an highly eccentric elliptical orbit around the hole, with an orbital period of roughly one hundred years. This is coupled to a complete reversal of the stellar spin axis: where at the beginning the star is characterized by a retrograde rotation, after the pericenter passage its rotation is prograde and still fairly intense ($\alpha\approx0.3$). 

These two facts bring to the conclusion that a century after the failed TDE there will be a regular (or possibly even an enhanced) one as the incoming star is, this second time, characterized by a rotation that is prograde with respect to the orbital motion.

\section*{Acknowledgements}
We ackowledge Chris Nixon and Ellen Golightly for interesting discussions and the anonymous referee for the insightful comments and suggestions.
All the snapshots of our simulations were obtained using SPLASH \citep{pri07}, an SPH data visualisation tool.




\bibliographystyle{mnras}
\bibliography{biblio} 

%




\bsp	
\label{lastpage}
\end{document}